%% file: icimp09-1fig.tex
\documentclass[times, 10pt,twocolumn]{article}

\usepackage{latex8}
\usepackage{times}

\usepackage{graphicx}
\graphicspath{{Figures/}}
\usepackage{url}

\newcommand{\ie}{{\em i.e.}}
\newcommand{\traceroute}{{\tt trace\-route}}
\newcommand{\tracetree}{{\tt trace\-tree}}
\newcommand{\scaleclem}{0.45}

\newcommand{\noteperso}[1]{\begin{center}
\renewcommand{\noteperso}[1]{}
\fbox{\begin{minipage}{6cm}#1\end{minipage}}\end{center}}

\begin{document}

\title{Fast dynamics in Internet topology: \\ observations and first explanations}
\author{Cl\'emence Magnien$\mbox{}^{1,2}$,
 Fr\'ed\'eric Ou\'edraogo$\mbox{}^{1,2,3}$, Guillaume Valadon$\mbox{}^{1,2}$, Matthieu Latapy$\mbox{}^{1,2}$\\
1: UPMC Univ Paris 06, UMR 7606, LIP6, F-75016, Paris, France\\
2: CNRS, UMR 7606, LIP6, F-75016, Paris, France\\
3: University of Ouagadougou, LTIC, Ouagadougou, Burkina Faso
\email{First-name.Last(-)name@lip6.fr}}

\maketitle{}

\thispagestyle{empty}

\begin{abstract}
By focusing on what can be observed by running \traceroute{}-like measurements
at a high frequency from a single monitor to a fixed destination set, we show
that the observed view of the topology is constantly evolving at a pace much
higher than expected.  Repeated measurements discover new IP addresses at a
constant rate, for long period of times (up to several months).

In order to provide explanations, we study this phenomenon both at the IP, and
at the Autonomous System levels.  We show that this renewal of IP addresses is
partially caused by a BGP routing dynamics, altering paths between existing
ASes. Furthermore, we conjecture that an intra AS routing dynamics is another
cause of this phenomenon.
\end{abstract}

\section{Introduction}

\input{intro}

\section{Data set}
\label{sec_data}

We use the data described in~\cite{radar}.
Measurements were conducted from more than 150 monitors.
Each monitor had a {\em destination set} that stayed
the same for the whole duration of the measurements.
The measurements then consisted in periodically running the  \tracetree{}
tool,
which collects a routing tree from a given monitor to a
set of destinations in a \traceroute{}-like manner.
The measurements were conducted with a high frequency
(typically about $100$ measurement rounds per day),
for a long period of time (from weeks to several months, depending on the monitor).
For 
more details, see~\cite{radar}.

Our goal is not to study the data in detail or compare all the  data sets obtained
from all monitors.  On the contrary, we insist on the fact that we observed
{\em similar} phenomena for each of them: while the exact details do of course
depend on the particular monitor under study, our observations were {\em
qualitatively} the same in all cases.

In this paper, 
we have therefore chosen to illustrate our results by using a single monitor
and the corresponding data set.  It consists of a single two-month measurement
(June and July $2007$) from a monitor located in Japan, at a rate of
approximately 100 rounds per day, leading to $5\,891$ rounds in total.  The
destination set consists of $3\,000$ IP addresses chosen randomly that replied
to {\tt ICMP echo request} messages at selection time.

\section{IP addresses renewal}

\begin{figure}
\begin{center}
\includegraphics[scale=\scaleclem]{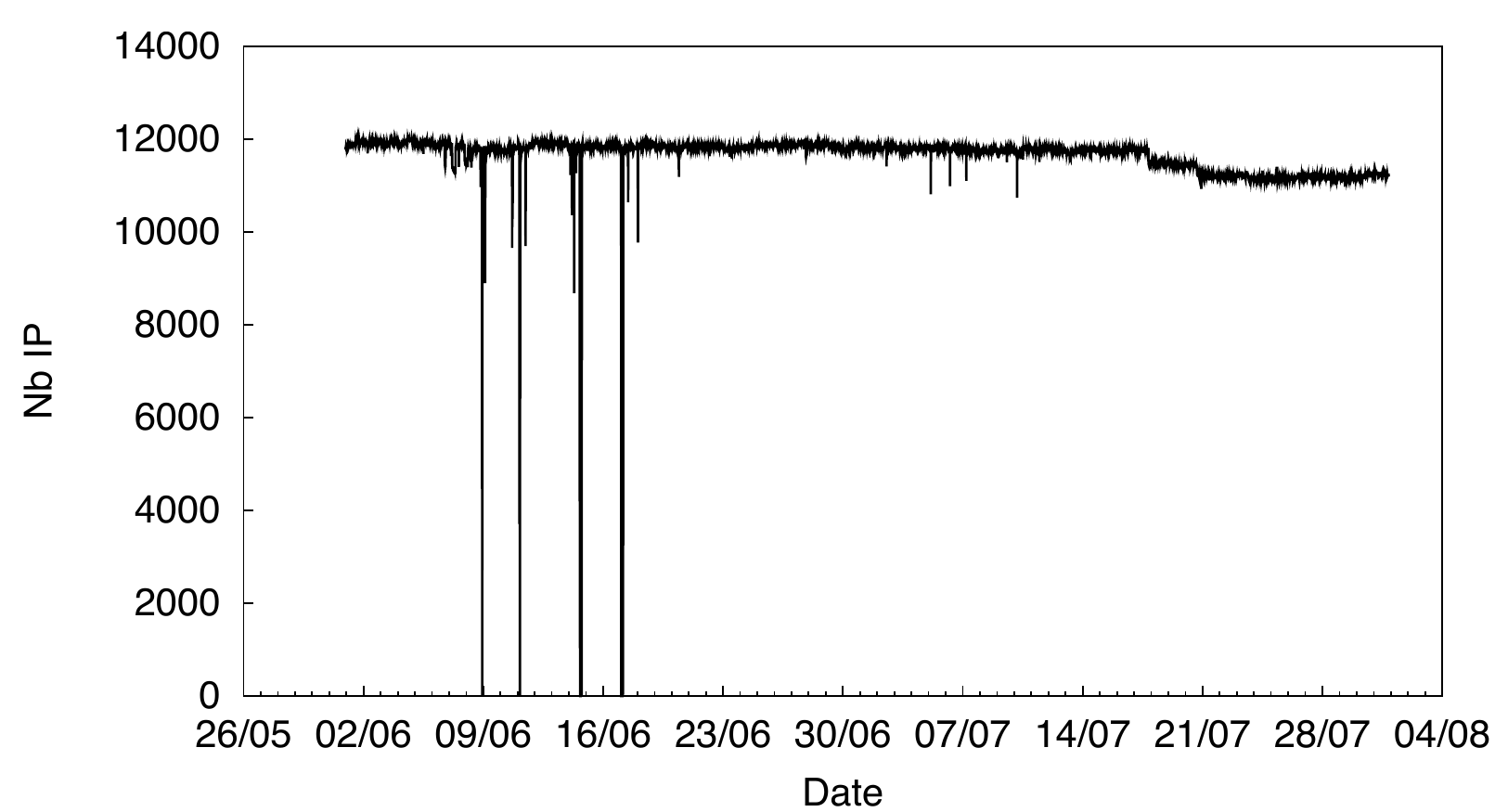}
\caption{Number of IP addresses observed in each measurement round.}
\label{fig_japon.nb_ip}
\vspace{-8mm}
\end{center}
\end{figure}

In this section, we describe  the evolution of  the set of observed IP
addresses.  Figure~\ref{fig_japon.nb_ip} presents the number of IP addresses
observed in each measurement round.  All values in this plot are centered
around a same value (close to $12\,000$) except some downward peaks which
indicate rounds  with {\em less} IP addresses than usual.  These peaks could
indicate a  loss of connectivity at or near the monitor, or an event such as a
major routing change or failure.  Studying this is however out of the scope of
this paper.

\begin{figure}
\begin{center}
\includegraphics[scale=\scaleclem]{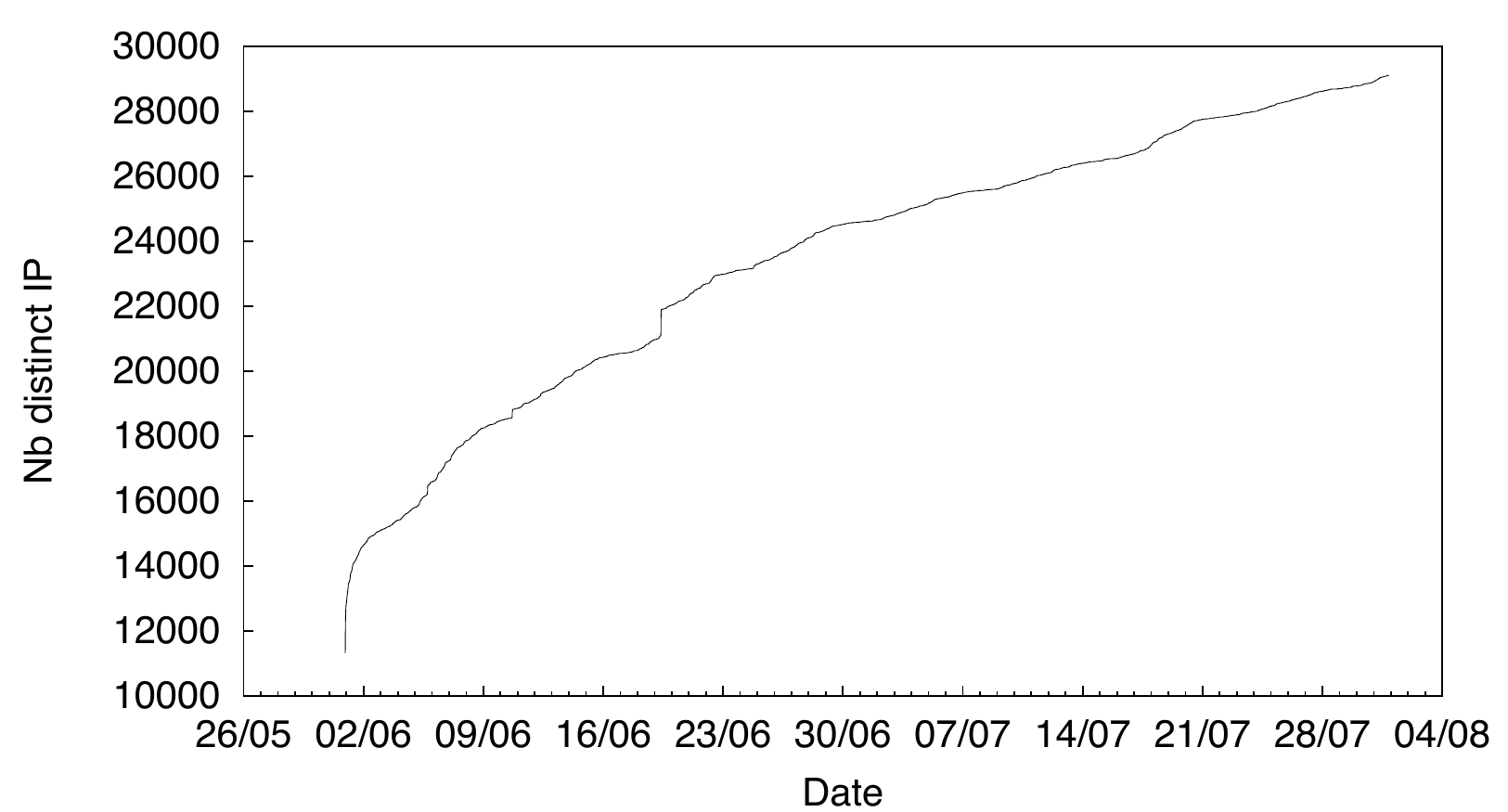}
\caption{
Number of IP addresses observed  since the beginning of the
measurements as a function of time. 
}
\label{fig_clem.nbs}
\vspace{-8mm}
\end{center}
\end{figure}

\begin{figure}
\begin{center}
\includegraphics[scale=\scaleclem]{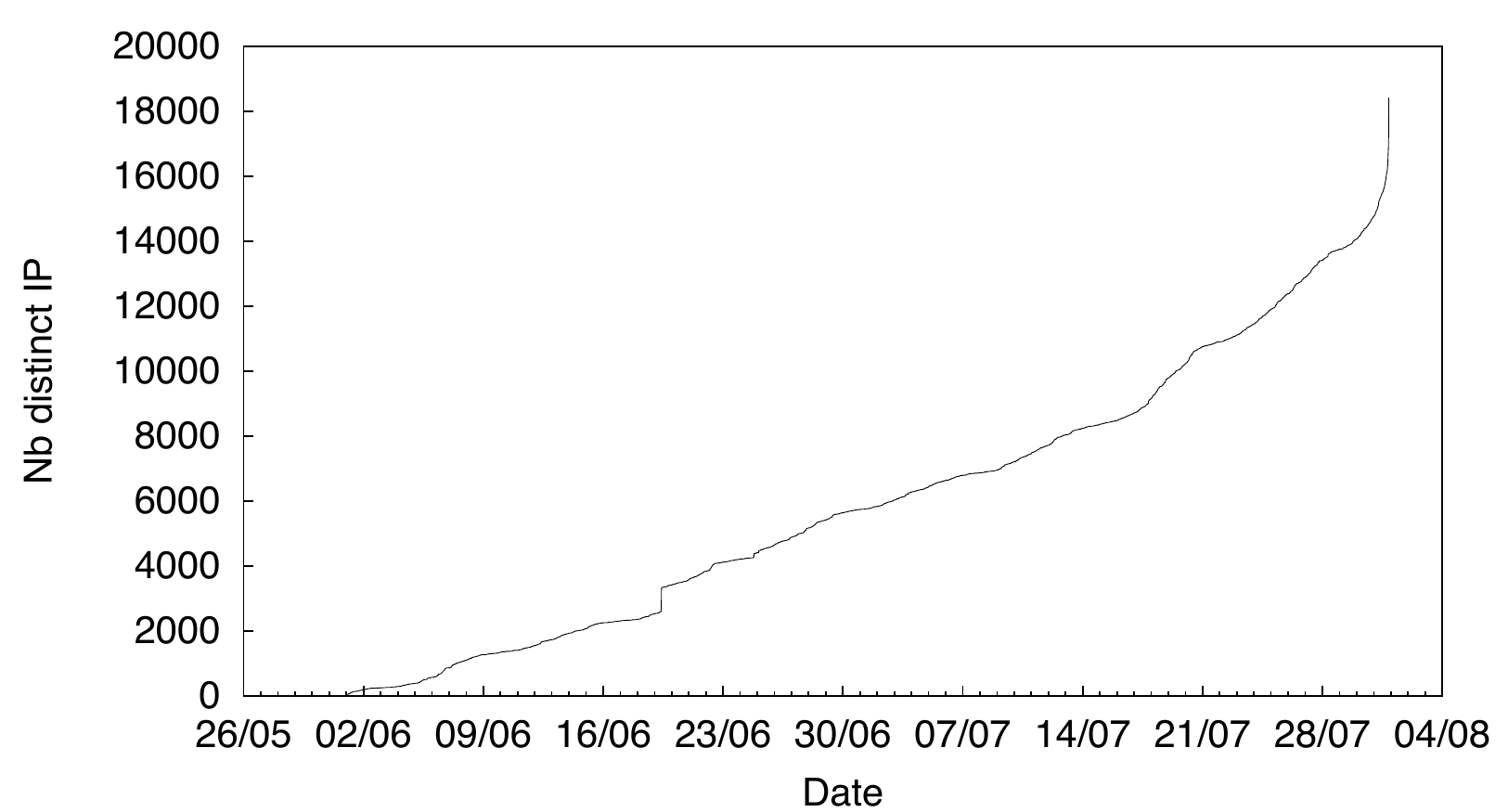}
\caption{
Number of IP addresses that were
observed before time $t$ and are never observed after $t$ as a function of
$t$.}
\label{fig_clem.nbs_rev}
\vspace{-8mm}
\end{center}
\end{figure}

The next question is whether we observe the same IP addresses  in all rounds.
This leads to the plot in Figure~\ref{fig_clem.nbs}, which presents
the number of distinct IP addresses observed since the  beginning of the
measurements as a function of time.  This plot gives evidence for a striking
fact: measurements continuously discover new IP addresses never seen
before\,\footnote{Other, six-months long, measurements exhibit the same
behavior.}.

Though it seems natural to observe {\em some} new IP addresses after some
measurement time, this happens here with a surprisingly high rate:
during the second month of the measurements,
around $150$ new IP addresses are discovered {\em each day}.

We have seen (Figure~\ref{fig_japon.nb_ip}) that the number of IP addresses
seen at each round is not increasing.  The continuous  discovery of new IP
addresses must therefore come together with a continuous  disappearance of
addresses that we cease to observe after some time.
Figure~\ref{fig_clem.nbs_rev} 
presents this.  The disappearances are
indeed symmetric with the observation of new IP addresses\,\footnote{The plots
of 
Figures~\ref{fig_clem.nbs} and~\ref{fig_clem.nbs_rev}
have a similar shape if
rotated at a $180^\circ$ angle.}.


One possible cause for these observations would be that some routers reply with  
random IP
addresses; we will show in the  
next
section that it is not the case.  Another possible cause would be that  
some of
our destinations are dynamic addresses, \ie{}
dynamically allocated to different hosts over time. Since such hosts  
could be
in different locations, depending on network operation, these  
dynamic
addresses could lead us to discover new paths and as a result new  
addresses
in the measurements.


\begin{figure}
\begin{center}
\includegraphics[scale=\scaleclem]{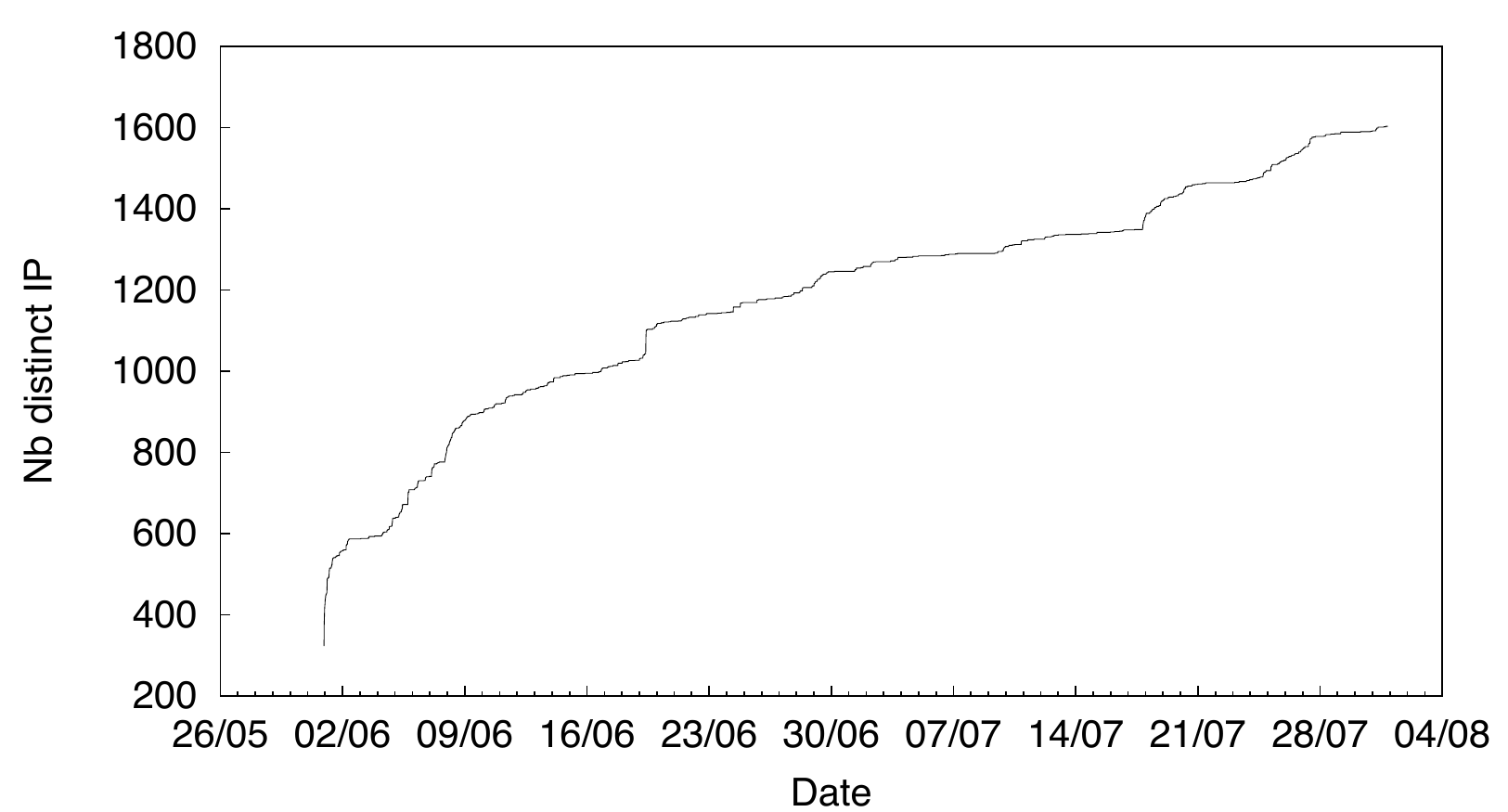}
\caption{Number of distinct IP addresses seen with stable destinations only.}
\label{fig_nbs_stable}
\vspace{-8mm}
\end{center}
\end{figure}

Figure~\ref{fig_nbs_stable} shows that this is not the case.  The idea is to
select the destinations that were {\em stable during the measurements}.  Using
a similar approach to geolocation studies (see for instance
\cite{PadmanabhanS01}), we considered that a destination address is not
dynamic if the address immediately before it in the measurements is always the
same; $35$ out of $3\,000$ destinations satisfied this condition.  We then
simulated the measurements by keeping only these stable addresses, and we
still clearly see a constant appearance of new IP addresses: dynamic addresses
are therefore not the cause of this renewal.  Note that our criterion for
characterizing stable addresses is very restrictive; we do not imply that the
addresses that do not satisfy it are dynamic.  We tested other criteria, which
provided the same results.

In summary, we observe a continuous, high-rate renewal of the set of IP
addresses observed from a monitor, and showed that it is not a measurement
artifact, but an actual property of the IP-level topology.  This implies that
repeating measurements, even for long periods of time, cannot converge to a
full view of what can be observed.  Moreover, aggregating data obtained during
consecutive rounds to
construct a topology map is not satisfying, because this means grouping
up-to-date data together with obsolete one.

\section{Recurring IP addresses}

\begin{figure}
\begin{center}
\includegraphics[scale=\scaleclem]{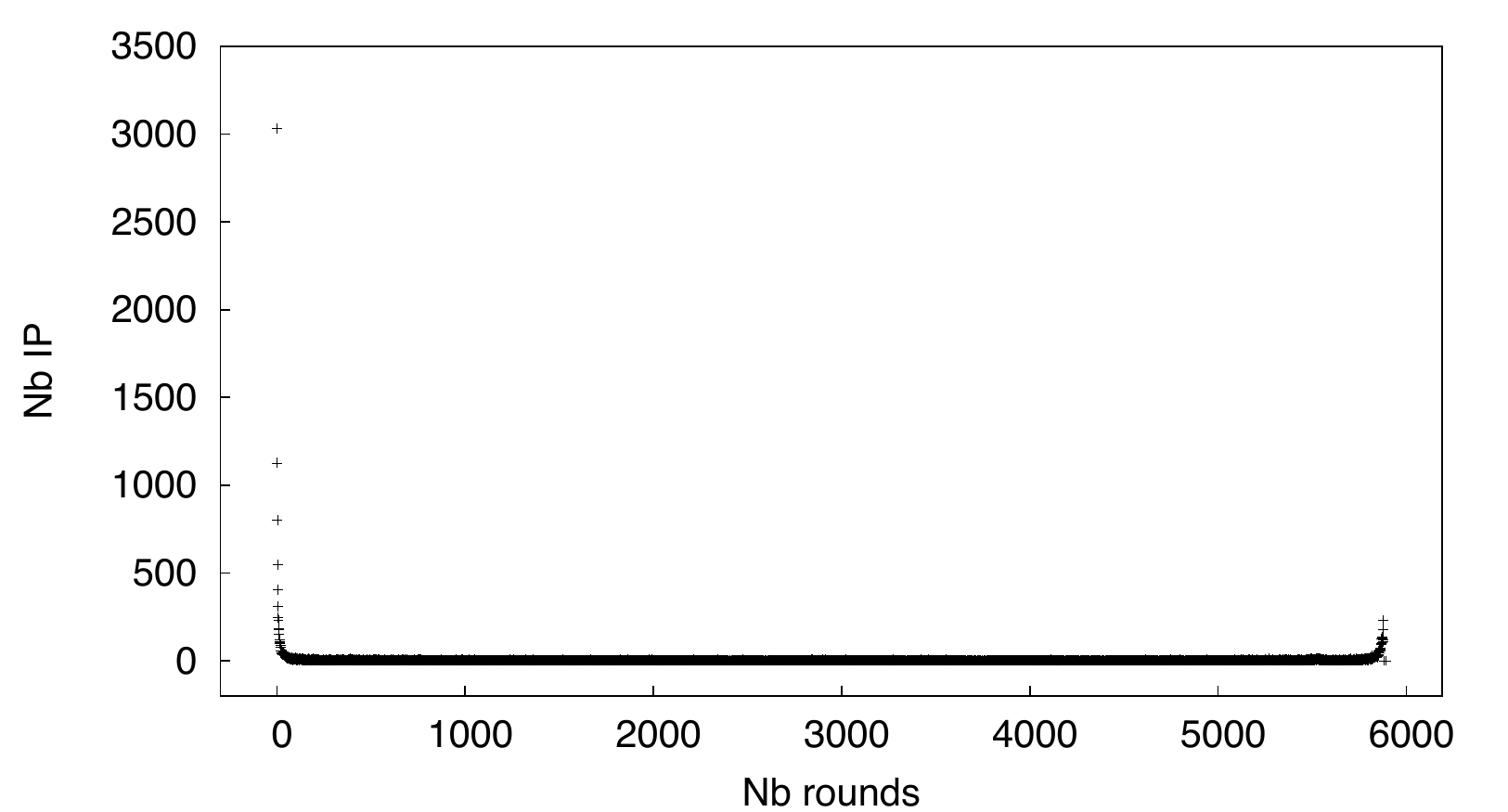}
\caption{
Distribution of the number of rounds in which each IP address
was observed. 
}
\label{fig_distrib}
\vspace{-8mm}
\end{center}
\end{figure}

\begin{figure}
\begin{center}
\includegraphics[scale=\scaleclem]{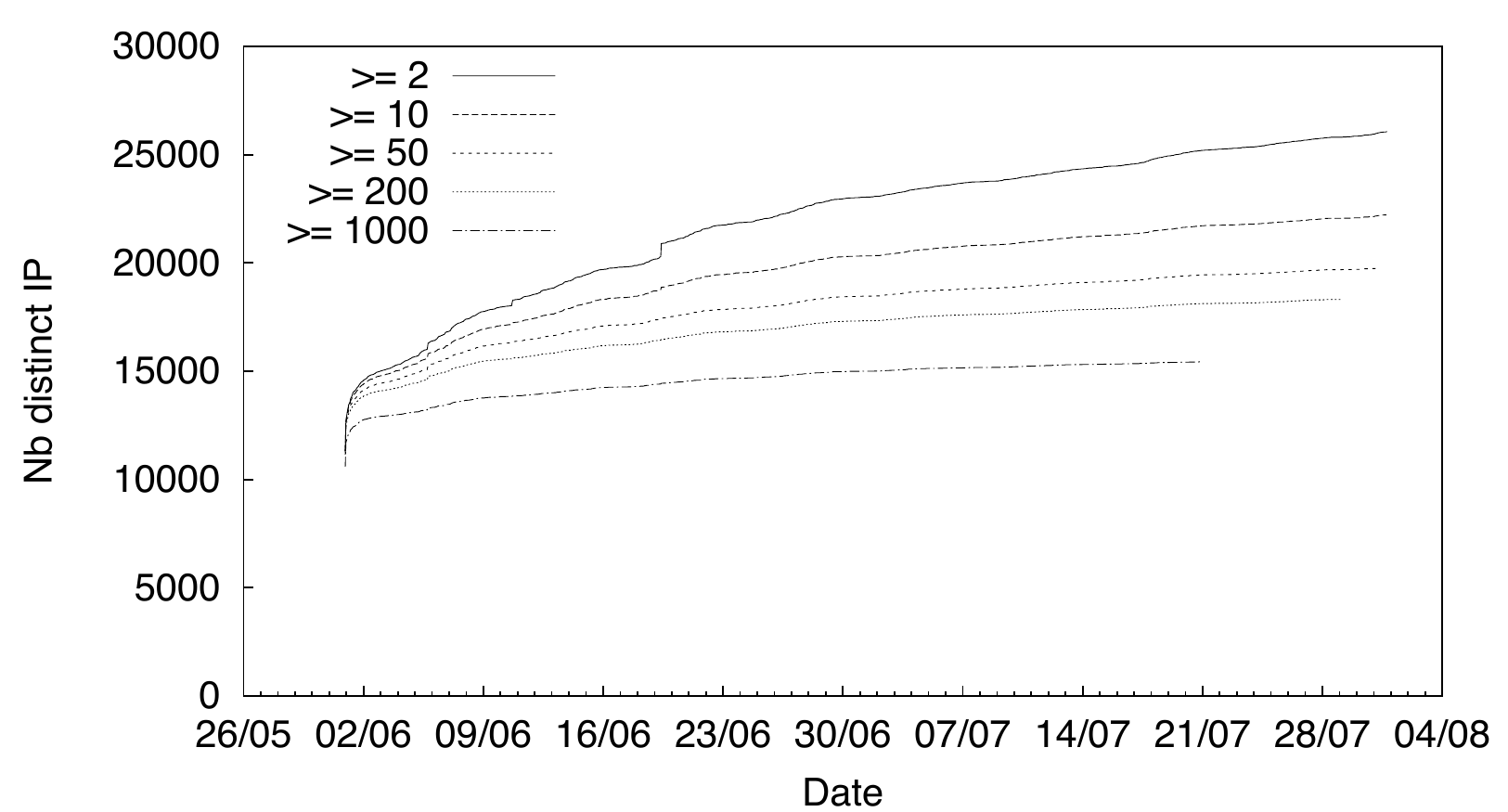}
\caption{
Number of IP addresses observed in at least 2, 10, 50,
200  or 1\,000 different rounds (top to bottom) since the beginning of the measurements.}
\label{fig_nbs_more1}
\vspace{-8mm}
\end{center}
\end{figure}

In this section we ask
whether
we observe IP addresses with consistency, or if we only see  
them in a
very small number of rounds.  
For any number of rounds $x$, Figure~\ref{fig_distrib} 
presents the number of IP addresses that  
were
observed in exactly $x$ different rounds during the measurements.

This distribution shows that a large number of IP addresses are very  
volatile:
$3\,030$ IP addresses are indeed observed only once during these two-month
measurements.  On the other hand, a significant number of IP addresses  
appear recurrently: they are seen in almost each round during the
measurements.


The presence of a large number of highly volatile IP addresses naturally
induces the question of whether these addresses are the cause of the
renewal
of the  observed IP addresses. Figure~\ref{fig_nbs_more1} 
answers
this question. It presents the number of distinct IP addresses
observed since
the beginning of the measurements, restricted to recurring addresses that were
observed in at least 2, 10, 50, 200 or $1\,000$ different rounds. Though
the slope of these plots are smaller than the one in Figure~\ref{fig_clem.nbs}, 
we continuously observe
new {\em recurring} addresses. As a matter of fact, if we only consider IP
addresses observed in least $1\,000$ rounds (out of $5\,891$ rounds), we
still
observe a non-negligible renewal. 
This shows that the constant observation of new IP addresses
is not caused by volatile  addresses only, and that
recurring addresses are also
renewed. Moreover, this means that
routers replying with random addresses are not the cause of this renewal: the corresponding
addresses would only be observed a very small number of times, and we
showed
that such addresses are not the main cause of our observations.

\section{Autonomous Systems}

\begin{figure}
\begin{center}
\includegraphics[scale=\scaleclem]{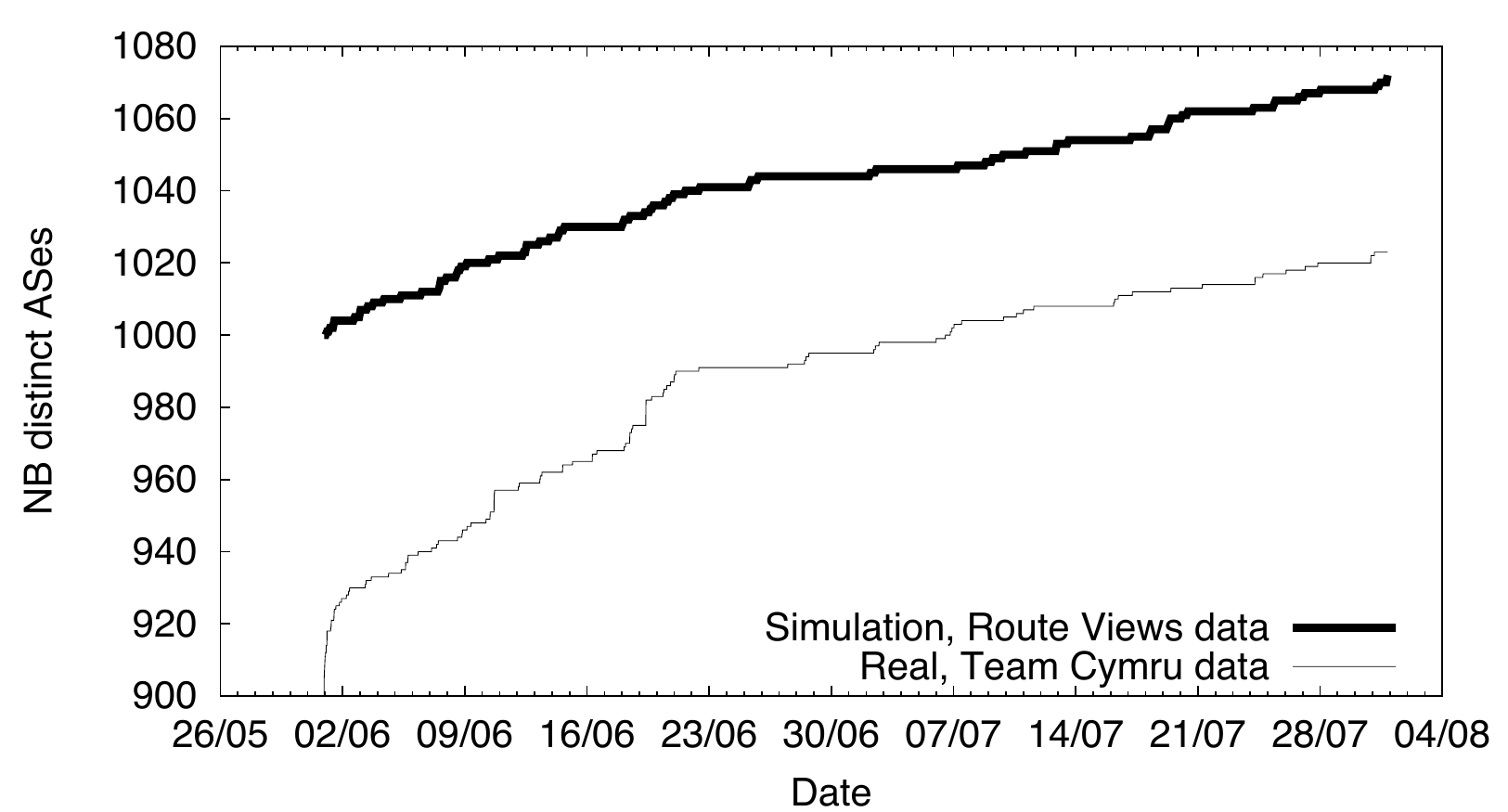}
\caption{
Number of distinct ASes observed since the beginning of  the
measurements: real number (thin line), and estimated with a simulation from
Route Views data (thick line).
}
\label{fig_nbs_as}
\vspace{-8mm}
\end{center}
\end{figure}

\begin{figure}
\begin{center}
\includegraphics[scale=\scaleclem]{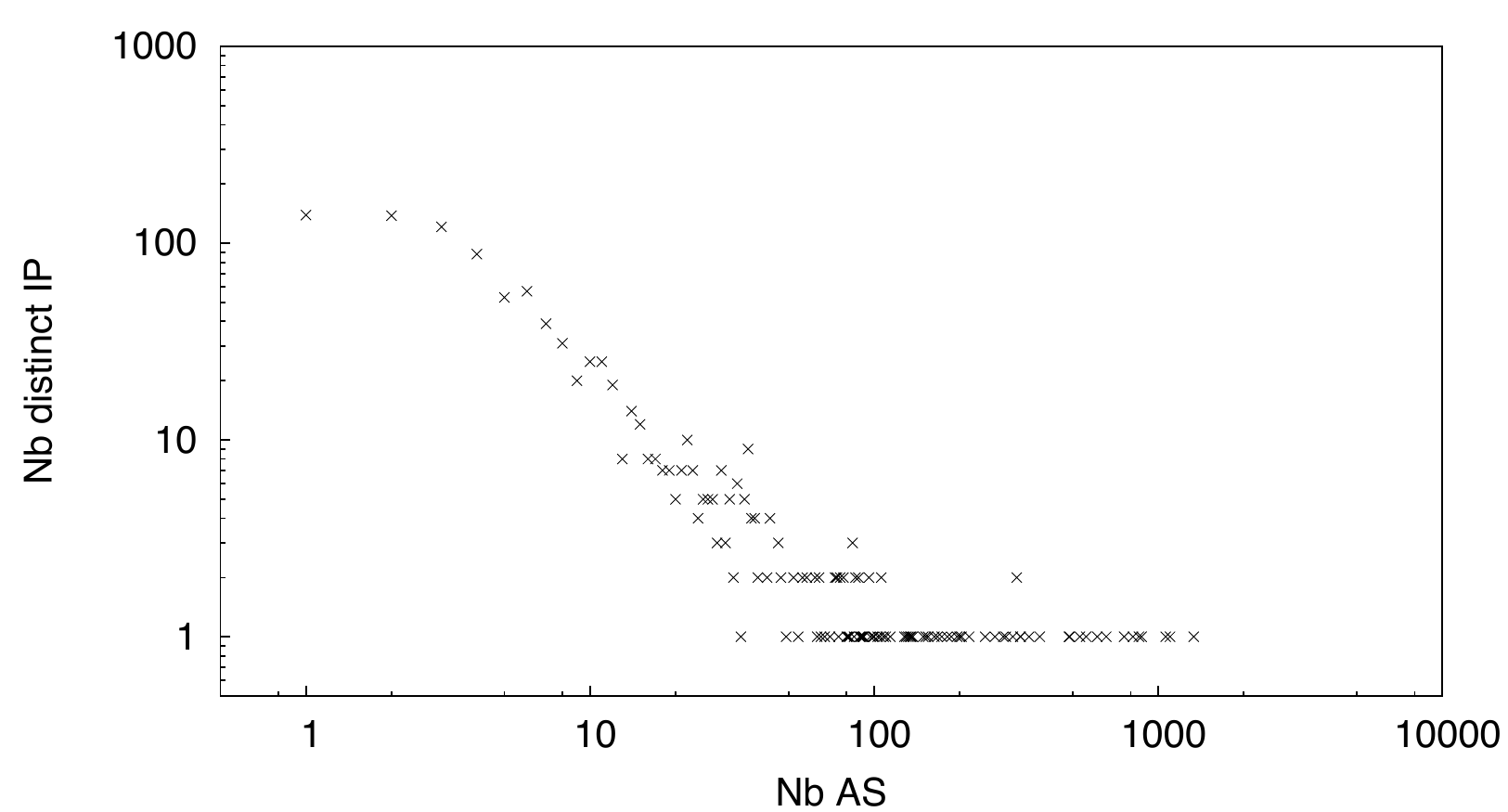}
\caption{
Distribution of the observed size of ASes.}
\label{fig_distrib.as}
\vspace{-8mm}
\end{center}
\end{figure}

We now study the same question on a different scale: 
do we observe the same type of behavior when we consider  
ASes rather than IP addresses?
We associate each IP address seen in the measurements to its AS  using the
Team Cymru database~\cite{cymru}.
Figure~\ref{fig_nbs_as} 
(thin line) then presents the number of distinct ASes
observed during the measurements.  As expected, we see fewer  
ASes than
IP addresses.  However, we observe the same {\em behavior} as before:  
 each round
sees a more or less constant number of ASes  (close to 950, not shown here),
and we
continuously discover new ASes during the measurements.

This can be considered as a partial explanation of what  
we
observe at the IP level: if we discover new ASes, it is only natural  
that we
should discover new IP addresses within them.


\begin{figure}
\begin{center}
\includegraphics[scale=\scaleclem]{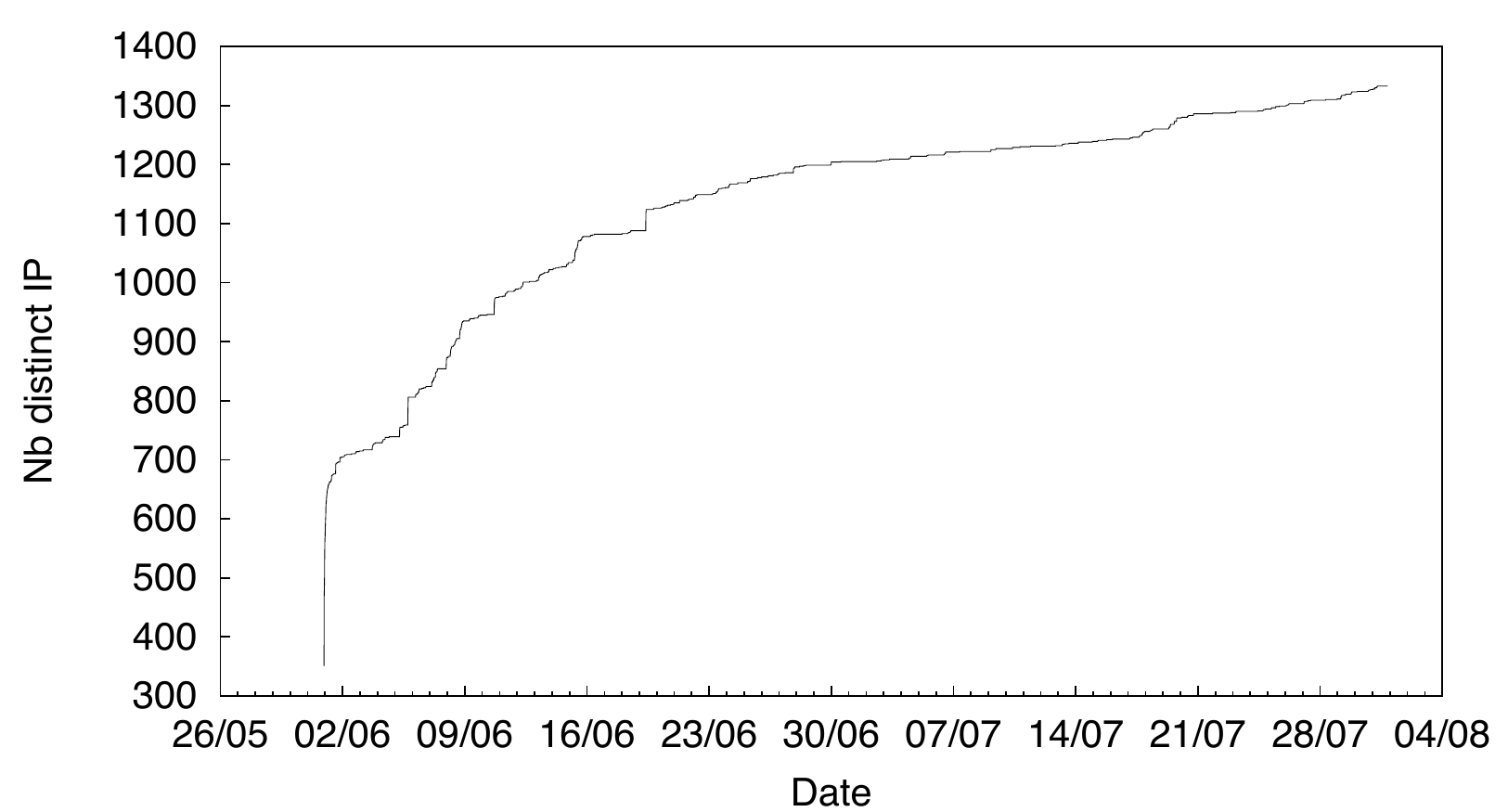}
\caption{Number of IP addresses observed in the
largest AS as a  function of time.}
\label{fig_as_3356}
\vspace{-8mm}
\end{center}
\end{figure}

To study this question further, we asked if all the ASes are  equivalent in
the measurements.  Figure~\ref{fig_distrib.as} 
shows the  distribution of the
observed size of ASes: for each AS seen, we computed how many  different IP
addresses we observed within it, and plotted the corresponding  distribution.
As we can see, this distribution is highly heterogeneous: for more than $100$
ASes (on a total of $1\,023$), we observe only a single IP address, while  3
ASes contain more than one thousand of observed IP addresses.

The presence of ASes with very large observed sizes in the measurements
naturally led to the question of whether  we also observe a renewal of
observed IP addresses {\em within} a single AS.  Figure~\ref{fig_as_3356}
shows the number of IP addresses observed since the beginning of the
measurements in the largest observed AS\,\footnote{This is the Level3
Communications AS (number $3\,356$), containing  $1\,333$ IP addresses (on a
total of $29\,100$).}.  Again in this  case we observe the same type of
behavior: we continuously observe new IP  addresses in this AS.

 From these observations we can derive the following conclusion: we  observe a
 constant renewal, both at the AS level, with the constant discovery of new
 ASes, and within single ASes, discovering new IP addresses in already seen
 ASes.   This therefore allows us to break down the constant appearance of new
 IP  addresses between these two factors.


However, this does not  explain {\em why} we should  
observe
new ASes, or new IP addresses within previously observed ASes.  In  
particular,
the question that naturally arises  about the newly discovered ASes is  
whether
they are {\em new}, \ie{} created after the beginning of the
measurements\,\footnote{During the year 2007, around 250 ASes were  
created every
month, see \url{http://www.cidr-report.org/as2.0/}}, or if they are pre-existing ASes that become
visible to the measurements due to BGP routing dynamics.

To study this question further, we used data from the Route Views
project~\cite{routeviews}.  This project makes publicly available a  recording
of BGP routing tables from several hosts.  This data allowed us to  
simulate
the measurements from an AS/BGP point of view: we chose a Route Views  
monitor
located close to our monitor\,\footnote{This is host
{\url route-views.wide.routeviews.org}, located in one peering point of the  
AS where
our monitor is located.}, then selected the routing  
tables
corresponding to the period of the measurements.
For each routing   table, we
extracted the
ASes belonging to BGP paths corresponding to IP  
prefixes of
the destinations. We thus obtained the set of all possible {\em observable}  
ASes for
each routing table.

Figure~\ref{fig_nbs_as} 
(thick line) then presents the number of
distinct  observable ASes since the beginning of the measurements, obtained
through our simulations.  We obtain a similar slope with both methods, which
confirms their validity.  Note that the numbers obtained with the Route Views
data are larger than  the ones obtained from Team Cymru.  This is due to the
fact that the Route  Views data represents, at each moment, the set of {\em
all} possible AS paths allowing to reach the destinations; instead, the Team
Cymru data  is directly extracted from the measurements, and therefore
provides only a single path to each destination.

The use of the Route Views data moreover allows us to go further. We  
observed
$1\,072$ ASes in total, 72 of which were discovered after the beginning of  
the
measurements.  Out of these ASes, we found out that 70, \ie{} all  
but two of them, were present in the {\em first} routing table
(but did not belong to AS paths leading to the destinations). 
This means  
that these
70 ASes were already existing at the beginning of the measurements,  
but became
visible because of BGP routing changes.

Finally, we are able to conclude that, at the AS level,
our observations are caused by a dynamics of the BGP routing,
causing pre-existing ASes to become visible on the paths between the
monitor and the destinations.

\section{Related work}

\input{related}

\section{Conclusion and perspectives}

\input{conclu}

\addcontentsline{toc}{special}{References}
\bibliographystyle{latex8}
\bibliography{biblio}

\end{document}

%% file: intro.tex
Most works aimed at mapping the Internet IP-level topology
rely on \traceroute{}-like probes,
for instance \cite{skitter,dimes}.
These probes are repeated periodically for large amounts of time,
each round of measurement leading to a partial and biased view
of the topology.
It is indeed known that, 
because of phenomena such as load balancing~\cite{parisTraceroute}, 
it is not possible to see everything that can
be seen from a monitor in a single round.
One round discovers only one path among several between the monitor and a destination.
Snapshots of the Internet topology are therefore constructed by merging series of
measurement rounds. 
This relies on the assumption that 
it is possible to explore a given part of the topology with a finite number
of probes.

We focus here on what can be observed by running \traceroute{}-like
probes at a high frequency from a single monitor to a constant
destination set~\cite{radar}.
We show that the observed view of the topology is constantly evolving
at a rate much higher than expected.
For instance, during the last week of two-months measurements,
we discovered $1\,118$ new IP addresses (on a total of $29\,100$)
that had never been observed before.

These observations imply in particular that
it is never possible to discover everything that can be seen from a monitor;
also, aggregating data from such measurements leads to topology maps with much obsolete information.

In this paper we describe and study this phenomenon.
Though we do not obtain a conclusive explanation,
we show that a fast routing dynamics is the cause.






%% file: related.tex
Much work has focused on the measurement bias created by mapping the Internet
topology with \traceroute{}-like probes.  The majority of these works concern
the fact that running probes from a limited number of monitors misses some
links and/or creates a bias on the observed degrees of the nodes, see for
instance~\cite{crovella,clausetmoore,plrevisited,spring2002measuring,guillaume2005metro}.
Others have studied the fact that tools such as \traceroute{} may report
incomplete and/or false information, see for instance~\cite{parisTraceroute,moors2004streamlining,paxson1999endtoend,wang2006quant,huffaker2002topology,teixeira2003search,spring2002measuring}.

It is an acknowledged problem
in the field
that the Internet topology evolves with time
and that this may create a bias in the measurements.
However, though some works have studied the dynamics of the
topology, at the IP or AS level
(see for
instance~\cite{plrevisited,paxson1999endtoend,wang2006quant,pansiot2007multicast,oliveira2007evol,park2004topology,park03static,stability,LabovitzMJ99,lad2006viulizing}),
up to our knowledge only one paper has attempted to study the bias caused
by this dynamics on the measurements~\cite{oliveira2007evol}.
The authors of this paper study the AS-level topology,
and design methods for evaluating with a certain degree of confidence
if an {\em observed} topology change is a {\em real} change or not.
Though their approach and some of their observations are similar to ours,
they study the AS-level topology whereas we study the IP-level topology,
and consider time-scales much longer, and hence a much coarser time resolution,
than we do.
The phenomena playing a role in their observations are therefore different
than in our case: 
they decompose their observations into a birth/death process,
coupled with transient routing dynamics.


Finally, another work~\cite{measurement} studied the measurement process of
different complex networks, including Internet maps.
They observed that, for the skitter data~\cite{skitter},
measurements continuously discover new IP addresses.
This is similar to our observations,
though other causes probably play a role in this:
the skitter data is collected from several monitors,
at a lower frequency and for larger time scales than the data we study here.




%% file: conclu.tex
In this paper, we bring to light a surprising phenomenon: when performing
periodic \traceroute{}-like measurements from a single monitor to a fixed set
of destinations, the obtained view of the topology never stabilizes.  On the
contrary, we continuously observe new IP addresses at a rate much higher than
expected.  This phenomenon is observed with various monitors and destination
sets, and seems to be universal.

We described this phenomenon in details and attempted to determine its cause.
We first ruled out some possible explanations:
dynamic IP addresses among the destinations,
and routers answering with many random addresses.
This showed that this phenomenon is not a measurement artifact,
but an actual property of the IP-level topology.

We were able to break down the observation of new IP addresses into
two factors:
a constant observation of new ASes, 
coupled with an observation of new IP addresses within already
discovered ASes.

From the record of routing tables by the Route Views project,
we concluded that the discovery of new ASes
is caused by the dynamics of the BGP routing.
These ASes 
were in fact created before the beginning of the measurements, and became visible 
as a consequence of a change in routing paths.


\medskip
Following these conclusions on AS renewal, we conjecture that the same cause
holds for the IP address renewal,
and that newly discovered addresses were already allocated at the beginning of the measurements,
and became visible because of routing changes.

\begin{figure}
\begin{center}
\includegraphics[scale=\scaleclem]{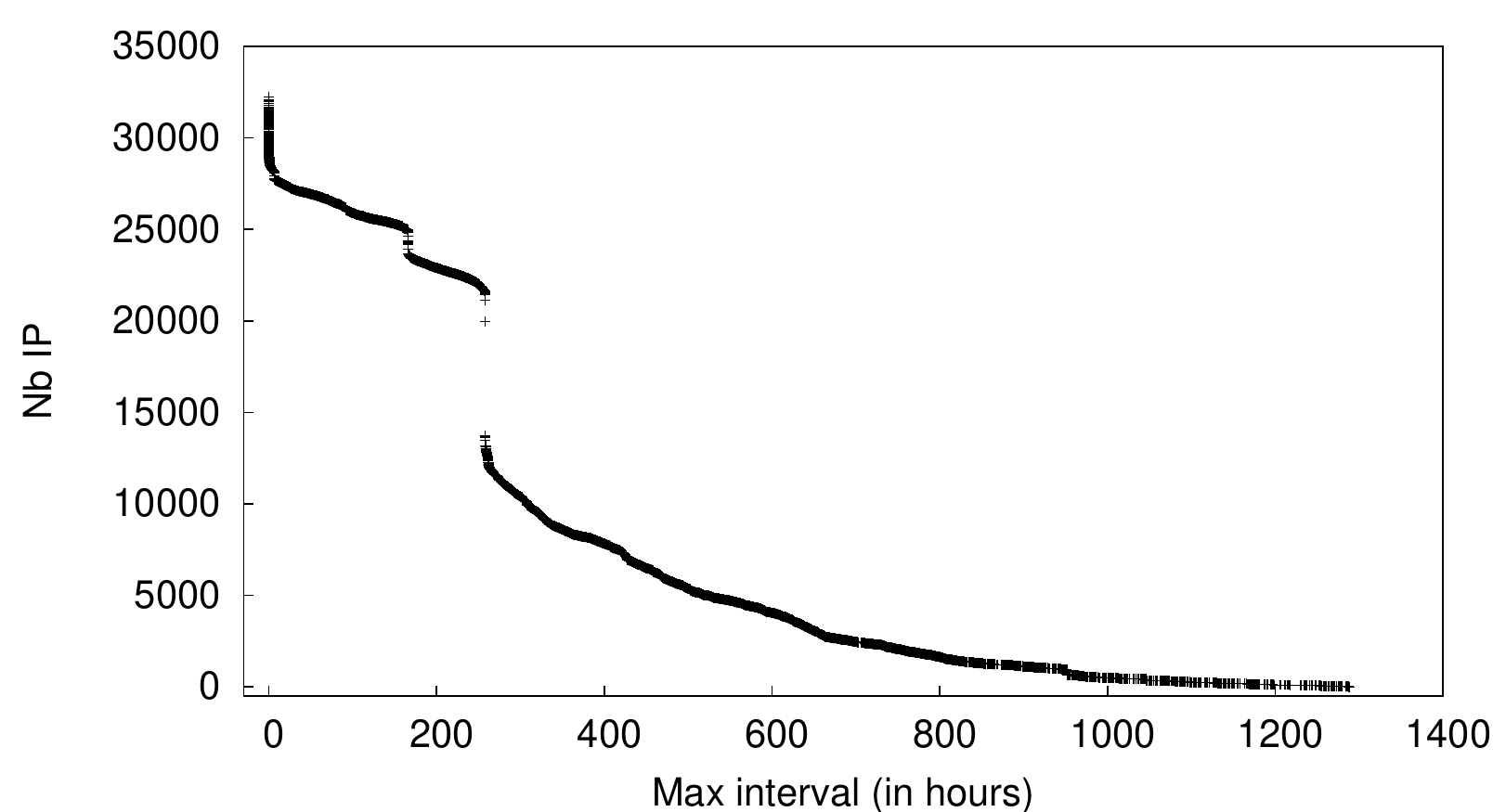}
\end{center}
\caption{Complementary cumulative distribution of the time elapsed between the first and last discovery of IP addresses
over all monitors.}
\label{fig_distrib_discovery}
\end{figure}

Some preliminary results on this question confirm this hypothesis.  We
combined measurements performed from several monitors in order to test if all
monitors discover the new IP addresses at the same time.  We chose $11$ monitors
that used the same destination set, 
and studied addresses observed with two monitors or more.
For each such address, we wrote down the time it was discovered
by each monitor individually and then computed the interval between the first
and the last of these discoveries.  For instance, an address seen with
two monitors, first observed with the first monitor at $7$ AM and then discovered
by the second monitor at $11$ AM the same day, will give an interval of four
hours.

Figure~\ref{fig_distrib_discovery} presents the complementary cumulative distribution of these interval sizes.
We observe that a large number of IP addresses discovered by a given
monitor were in fact observed a significant duration before with other monitors.
Among the $32\,228$ (out of $40\,076$) IP addresses seen with at least two monitors,
$22\,897$
were observed with one monitor more than $200$ hours  before they were discovered by another one,
which means that these addresses existed for a long time before they were
discovered.
Note that this does not tell us whether other addresses existed before their discovery
or were created at this time.
This indicates that a large number of the IP addresses discovered by a given monitor
existed in fact for a significant time before their discovery,
and that a routing dynamics between existing addresses plays a strong
role in our observations.

\medskip
This work should be pursued in several directions.
First, we want to fully characterize and understand the renewal of 
observed IP addresses.
We think that it is possible to perform new measurements,
specifically designed for answering the question of whether newly discovered
IP addresses existed prior to their discovery or not.
Another direction  would be to
perform simulations,
which would open the way to an accurate modeling of the phenomena 
causing the renewal of IP addresses.

Second, our work indicates that there is no perfect solution for constructing
maps of the Internet topology 
while a single measurement round does not discover everything that can
be seen from a monitor (because of load balancing for instance),
aggregating data from several consecutive rounds puts together obsolete
and up-to-date data.
It would be of prime interest to determine if  a best compromise exists,
allowing to construct the most accurate maps, for instance by tuning the number of measurement
rounds and their frequency.
